\documentclass[preprint,review,12pt]{elsarticle}

\usepackage[colorinlistoftodos]{todonotes}
\usepackage{placeins}

\usepackage{ecrc}

\volume{00}
\firstpage{1}
\journalname{Journal of Membrane Science}
\runauth{C. J. Linnartz et al.}
\jid{J. Memb. Sci.}
\jnltitlelogo{J. Memb. Sci.}

\usepackage{amssymb}
\usepackage{mhchem}
\usepackage{overpic}
\usepackage{graphics}
\usepackage{multirow}
\usepackage{rotating}
\usepackage{textcomp}
\usepackage{float}
\usepackage{color}
\usepackage{amstext}
\usepackage{siunitx}
\usepackage{subfigure}
\usepackage{graphicx,wrapfig,lipsum}
\usepackage{soul}
\usepackage{xr}
\usepackage{lineno, blindtext}

\begin{document}
\begin{frontmatter}

\title{Membrane-Electrode Assemblies for Flow-Electrode Capacitive Deionization}

\author[DWI,AVT.CVT]{Christian J. Linnartz}
\author[DWI,AVT.CVT]{Alexandra Rommerskirchen}
\author[AVT.CVT]{Joanna Walker}
\author[AVT.CVT]{Janis Plankermann-Hajduk}
\author[DWI,AVT.CVT]{Niklas K\"oller}
\author[DWI,AVT.CVT]{Matthias Wessling\corref{mycorrespondingauthor}}
\cortext[mycorrespondingauthor]{Phone: +49 241 80-95470, Fax: +49 241 80-92252, manuscripts.cvt@avt.rwth-aachen.de}
\address[DWI]{DWI - Leibniz-Institute for Interactive Materials e.V., Forckenbeckstra\ss e 50, 52074 Aachen, Germany}
\address[AVT.CVT]{RWTH Aachen University, Aachener Verfahrenstechnik-Chemical Process Engineering, Forckenbeckstra\ss e 51, 52074 Aachen, Germany}

\begin{abstract}

Scale-up of flow-electrode capacitive deionization is hindered due to the reliance on thick brittle graphite current collectors. Inspired by developments of electrochemical technologies we present the use of flexible membrane electrode assemblies (MEA) to solve these limitations.
We tested different carbon-fiber fabrics as current collectors and laminated them successfully with ion-exchange membranes. The use of thinner ion-exchange membranes is now possible due to the reinforcement with the carbon fiber fabric. 

Desalination experiments reveal that a MEA setup can achieve salt transfer rates equal to standard setups. Hence, we deduce that charge percolation also acts outside the electric field. In a single point of contact, ionic and electric charges are exchanged at the carbon surface of the MEA. The use of thinner membranes leads to a reduced potential drop. Together with a more homogeneous electric field across the feed water section, this can compensate for the reduction of contact surface between flow electrode and current collector.  

\end{abstract}


\end{frontmatter}


\section{Introduction}

The development of environmentally friendly and energetically efficient technologies for the desalination of water poses a major challenge on the way towards a sufficient and safe water supply. Additionally, many industrial processes produce salt-rich wastewater streams \cite{REDONDO2001}. Recovery and concentration of valuable ionic components from aqueous streams may improve the sustainability and economic feasibility of various large scale industrial processes. The implementation of new technologies for recycling of valuable ionic compounds from salt solutions reduces the water demand to prepare these solutions and minimizes the impact to the environment by not disposing them to the surface water body  \cite{OECD2019, Elimelech2011}. The approaches to solve these problems can be divided by the applied driving force. Besides pressure (reverse osmosis, nanofiltration) and heat driven processes (distillation, vapor compression, membrane distillation)  \cite{Amy2017,Phillip2011,Karagiannis2008}, electrochemical processes attain more and more attention since they allow the direct use of electricity from renewable resources and, thus, reduce carbon emissions \cite{SUBRAMANI2011}.  

In this field a vast variety of electrochemical processes evolved over the past years, such as electrodialysis (ED), which uses an electric field to move ions between different flows separated by ion-exchange membranes (IEMs). ED is used to set salinities in a large variety of applications; drinking water production is a major field, but it is also widely applied in the food industry. The technology is readily adaptable to the demands and can be scaled-up by stacking 50 to 100 cells, or layers of IEMs. However, the technology is not yet well applied at high salt concentrations and relies on faradaic reactions to couple ionic and electric charge transport. 

A more recent development is capacitive deionization (CDI), which is considered more suitable for the desalination of low concentrated salt waters. CDI also uses an electric field as driving force; the ions are removed from the solution and are stored in the electric double layer at the interface between solution and electrode via electrosorption. With the implementation of ion-exchange membranes, the electrical efficiency is improved, and the process was commercialised and applied for off-grid desalination \cite{Oren2008, Porada2013, Suss2015, Bales2019}.

A continuous version of CDI is flow-electrode capacitive deionization (FCDI), which is also applicable to high salt concentrations.\cite{Jeon2013, Gendel2014} Different strategies of charging and discharging the flow-electrodes were studied \cite{Rommerskirchen2015SingleDesalination, Porada-Mixing, Doornbusch-FluidizedFCDI} as well as a possible application enabeled by the unique separated transport of cations and anions in different flow electrodes\cite{Linnartz2017Flow-ElectrodeReactions}. Since the FCDI technology has many possible applications like salt metathesis and outperforms other technologies with its reduction of the energy demand \cite{Rommerskirchen2018EnergyProcesses, Linnartz2017Flow-ElectrodeReactions, Zhang2018}, scale-up strategies are studied and developed. Yang et. al. showed an FCDI Stack using graphite current collectors to charge the flow-electrodes \cite{Yang2016}. Cho et. al. continued with a segmented approach increasing the packing density of an FCDI cell \cite{Cho2017} using a ceramic support structure. Yang et. al. continued on our early cell designs by introducing an electrically conductive titanium mesh next to the ion-exchange membranes showing the reduction of the charge transfer distance \cite{YANG2019114904}. This work also leads to hybrids between FCDI and ED configurations \cite{MA2020115186}.  In this paper, we are taking an important step forward by introducing a membrane electrode assembly (MEA) to FCDI which uses low-cost, flexible and chemically stable carbon fiber fabric to conduct the electric charge and reinforce the ion-exchange membranes at the same time. Note that MEA essentially denotes the assembly of membrane and current collector.

\begin{figure}[h!]
\begin{center}
       \includegraphics[width=\textwidth]{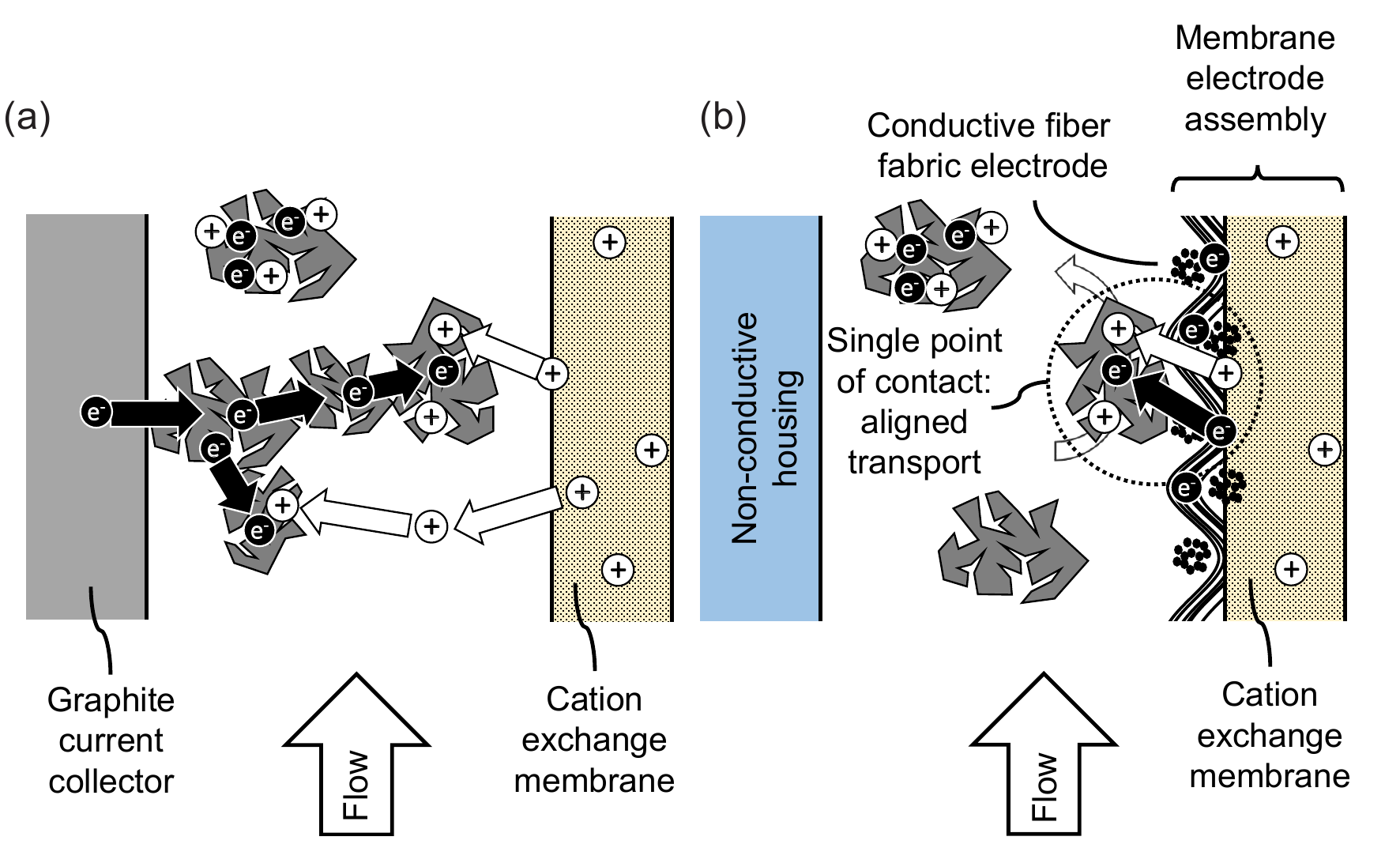}
\caption{Different explanation of charge transport between a surface and particles of a flow electrode. (a) Charge transport between graphite current collector and membrane \cite{Rommerskirchen2019, Hatzell2017}. Electrical charge transport by percolation of particles and ionic charge transport in the liquid phase. (b) Charge transport at the surface of a MEA. Direct exchange of electrical and ionic charges in a single point of contact.}
\label{fig:singlepoint}
\end{center}
\end{figure}

Earlier work always looked at charge transport from the conductive surface via the slurry electrode to the membrane as displayed in Figure~\ref{fig:singlepoint}a \cite{Rommerskirchen2019, Hatzell2017, Dennison2014}. Contrary, when using the proposed MEAs, only the surface facing the membrane is conductive. To describe this effect, we need a new approach to explain the charge transport in flow electrodes using MEAs. As shown in Figure~\ref{fig:singlepoint}b, our hypothesis is that the slurry particles exchange both, electrical and ionic charges, starting from a single point of contact. The electric charge is supplied by the carbon fiber fabric and ions which passed the ion-exchange membranes can adsorb to a flow-electrode particle at the same location. This is comparable to descriptions in literature, where particles are reported to take charge from a surface \cite{Rees2012}. In addition, charge percolation can still emerge forming percolation networks from the carbon fiber fabric towards the non-conductive flow-field. 

Apart from desalination, ion-exchange membranes and membrane electrode assemblies have a larger application in the field of electrolysis processes, fuel cells and batteries \cite{Baker.2012, PAIDAR2016}. In polymer electrolyte fuel cells (PEFCs), the front- and backside of the membranes are covered with a catalyst layer, where the electrochemical half cell reactions take place. On top of this catalyst layer, the gas diffusion layer (GDL) is placed, which ensures fast supply of the reactants to the catalyst layer and simultaneously acts as a current collector. Reactions in the gas phase are prominent since  diffusion coefficients are of magnitudes larger compared to a possible liquid phase. The materials applied for the GDL are commonly carbon based papers or woven fabrics, but also nickel and metal based alloy foams \cite{An2018, Gottesfeld2018, SupramaniamSrinivasan2006, Zhao201301, Merle2011}.

For proton exchange membrane fuel cells (PEMFC), membrane electrode assemblies are made by hot pressing. Suitable conditions have been identified to lie in the temperature range between 100-\SI{160}{\celsius} and a pressing time of 1-\SI{2.5}{min}. The applied pressures vary between 100-\SI{16000}{N/cm\textsuperscript{2}} and depend on the gas diffusion layer material and catalyst coating strategy. \cite{SupramaniamSrinivasan2006,Zhao201301,Yang_2003,Koziara.2016,thanasilp2010,Therdthianwong2007,Frey2004, YILDIRIM2008364}. The extreme forces on the membrane can cause unseen damage and change the membrane properties, e.g. by deterioration of functional groups. Therefore, the temperature and pressure should be chosen carefully. The ion-exchange capacity of the membranes before and after the hot-pressing procedure has been investigated, showing that their properties stay the same \cite{SupramaniamSrinivasan2006,roy2009}.

The hypothesis of this work is that by integrating the developments in designing a membrane electrode assembly (MEA) for fuel cell applications and the developments in FCDI, one can improve the overall process of FCDI. The aim is to achieve a reduction of resistances in the cell as well as a thinner and more flexible stack design. Therefore, we report a specific new MEA design by adaption of the production procedure from literature and demonstrate the implementation in an FCDI cell.

\section{Experimental}
\subsection{Manufacture of Membrane Electrode Assemblies}
\label{sec:MEAmanufac}

The membrane electrode assemblies (MEAs) we fabricated and tested in this study are sandwiched from ion-exchange membranes (FAS~30 and FKS~30, FUMATECH BWT GmbH) and different carbon-fiber fabrics (CF) without any additional material. For clarity, each carbon fiber fabric was abbreviated with CF and their specific surface weight.

\begin{table}[h!]
    \centering
    \begin{tabular}{c l}
         CF46 &  weight surface: 46~g/$m^2$, thickness: 60~\textmu m, thread diameter: 7~\textmu m \\ 
            & thread count: 7000, weave: plain, non-coated \\ 
            &provided by R \& G Faserverbundwerkstoffe GmbH, SAMURAI Carbon Fabric
            \\ \hline
            CF92 & weight surface: 92~g/$m^2$, thickness: 150~\textmu m, thread diameter: 6~\textmu m \\ 
            & thread count: 1000, weave: plain, non-coated\\ 
            & provided by Goodfellow GmbH, C003510  
            \\ \hline
            CF200 & weight surface: 200~g/$m^2$, thickness: 300~\textmu m, thread diameter: 7~\textmu m \\ 
            & thread count: 3000, weave: plain, non-coated \\ 
            & provided by Goodfellow GmbH, C003531
    \end{tabular}
    \caption{Different carbon fiber fabrics applied in this study.}
    \label{tab:DiffFibers}
\end{table}

We used a hot pressing procedure to join the single layers. Membrane and carbon fiber fabric were stacked and covered by a layer of aluminum foil for surface protection and placed in a hot press (Agila PE30, Belgium) ensuring an even distribution of heat and pressure. The structures were pressed for \SI{2.5}{min} at \SI{120}{\celsius} and \SI{8.5}{\bar}, immediately followed by a second \SI{2}{\min} pass at the same pressure but at a temperature of \SI{25}{\celsius}. 
To ensure the MEA has good sealing properties in an FCDI module, the side of the MEA with exposed carbon fiber fabric was sealed on the rim with polydimethylsiloxane (Sylgard\textsuperscript{\textregistered} 184 by Dowsil). The active area was spared out, so ions can pass through the membrane and fiber fabric and are stored in the flow electrode. The electrical contact was made by a copper pop rivet clenched into the fibers at an extension (see Figure~\ref{fig:SEMMEA}).

\subsection{FCDI modules}

\begin{figure}[h!]
	\centering
		\includegraphics[width=\textwidth]{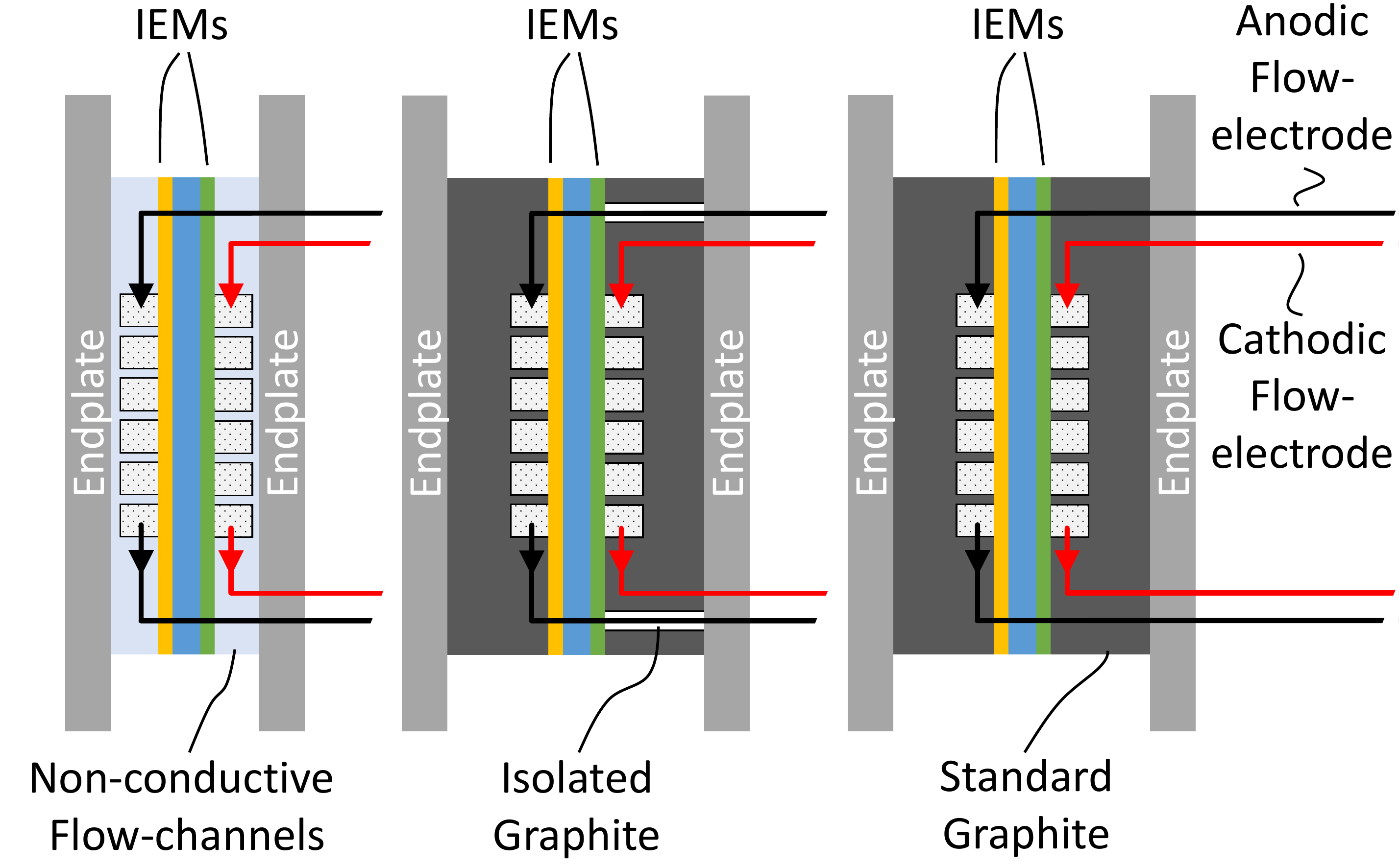}
	\caption{Different configurations of desalination modules. Left: Setup with MEAs and a non-conductive flow-channels. Center: Setup with graphite current collectors and isolated flow electrode bore holes. Right: Setup with standard graphite current collectors.}
	\label{fig:ModuleComparison}
\end{figure}

In the present study, two sizes of FCDI modules were used. They are termed as 'standard' and 'small' modules. The construction of the 'standard' module was almost identical to the module used in our previous works \cite{ Rommerskirchen2015SingleDesalination, Linnartz2017Flow-ElectrodeReactions,Gendel2014ElectrochemistryTechnology}, but no titanium fleeces were placed between the ion-exchange membranes and current collectors \cite{Rommerskirchen2018EnergyProcesses, Rommerskirchen2019}. Those modules were made from polymer end-plates, rubber gaskets, epoxy-impregnated graphite plates (M\"uller \& R\"ossner GmbH \& Co. KG, 180 $\times$ 180 $\times$ \SI{10}{mm^3}) with engraved flow-channels for the flow electrodes (3 $\times$ \SI{2}{mm^2} and 200~cm total length), Polyether ether ketone (PEEK) reinforced cation- and anion-exchange membranes (Fumasep FKB-PK-130/ED-100 and Fumasep FAB-PK-130/ED-100, FUMATECH BWT GmbH) with an effective area of \SI{121}{cm^2}, and a \SI{0.5}{mm} mesh spacer (ED-100 Spacer, FUMATECH BWT GmbH). The 'small' graphite modules were made from the same materials, but in smaller dimensions of polymer end-plates and rubber gaskets. We used epoxy-impregnated graphite current collectors (M\"uller \& R\"ossner GmbH \& Co. KG, 62 $\times$ 128 $\times$ \SI{10}{mm^3}) with engraved flow-channels for the flow electrodes (3 $\times$ 2~$mm^2$ and 470~mm total length), PEEK reinforced cation- and anion-exchange membranes (Fumasep FKB-PK-130/ED-40 and Fumasep FAB-PK-130/ED-40, FUMATECH BWT GmbH) with an effective area of \SI{22.4}{cm^2}, and a \SI{0.5}{mm} mesh spacer (ED-40 Spacer, FUMATECH BWT GmbH) for the reference experiments. To study short circuiting effects, we also integrated polyethylene tubing as isolation in a module with graphite current collectors. For the MEA experiments the same flow-channel geometry as in the graphite current collectors of the 'small' modules was engraved to polycarbonate plates, and the MEAs were used instead of the aforementioned ion-exchange membranes. The MEAs were installed such that the carbon fiber fabric always faced the polycarbonate plates. The three different setups are shown in Figure \ref{fig:ModuleComparison}.

\subsection{Ion-Exchange Capacity}

The influence of the proposed heat pressing procedure on the membrane functionality was evaluated by the ion-exchange capacity (IEC). For determination of the IEC, the procedure proposed by Sata and Dlugolecki was applied.  In \mbox{Equation \ref{eqn:ion exchange capacity}} the ion-exchange capacity is defined as total amount of charged groups \mbox{c\textsubscript{total} [mequiv]} over dry weight \mbox{m\textsubscript{dry} [g]} of the membrane \cite{Dlugolecki2008, Sata2007}.
\begin{equation}\label{eqn:ion exchange capacity}
    \mathrm{IEC} = \frac{\mathrm{c}\textsubscript{total}}{\mathrm{m}\textsubscript{dry}}
\end{equation}
As it is not possible to measure these groups directly, the amount of corresponding ions that are replaced by other ions is determined. Therefore, the membranes are first brought into a specific form, meaning functional groups are occupied by one species of ion. Prior to this, the exact dry weight of the investigated piece of membrane has to be measured.

Cation-exchange membranes were brought into H\textsuperscript{+}-form by immersing them into \SI{1}{\mol/L} hydrogen chloride solution (Carl Roth GmbH \& Co. KG) for at least 24h, replacing the solution three times. Afterwards, the membranes were thoroughly rinsed with ultra-pure water to remove any remaining protons on the membrane surface. The rinsing water was tested for remaining chloride ions by addition of silver nitrate. Subsequently, the membranes were brought into Na\textsuperscript{+}-form by laying them into \SI{3}{\mol/L} sodium chloride solution (Carl Roth GmbH \& Co. KG, 99.8~$\%$) under continuous stirring. To ensure complete exchange of protons with sodium ions, the sodium chloride solution was replaced three times within \SI{24}{h}. The replaced solution was collected and its proton concentration was determined by acid-base titration.

Anion-exchange membranes were brought into Cl\textsuperscript{-}-form by soaking them into \SI{3}{\mol/L} sodium chloride solution for at least 24 h during which the solution is replaced three times. Further, the membrane was rinsed with ultrapure water to remove adsorbed chloride ions. Again, the rinsing water was tested on remaining chloride ions by addition of silver nitrate. The membrane, in balance with ultrapure water was then leached in \SI{0,2}{\mol/L} sodium nitrate solution (Carl Roth GmbH \& Co. KG, purity $\geq$ 99\,\%) under continuous stirring. The solution was replaced three times to ensure complete exchange of chloride ions through nitrate ions. Again, the replaced solution was collected and its chloride content was determined by high performance liquid chromatography (Agilent 1200, HPLC).

\subsection{Electrochemical Impedance Spectroscopy}

Electrochemical impedance spectroscopy was performed on the set of membranes and membrane electrode assemblies applied in this study to get an insight into the resistance and double layer formation on the membrane surfaces. A four electrode setup from Luo was used with a two compartment tubular test cell \cite{Luo2017}. The surface under examination had a size of \SI{1,77}{cm\textsuperscript{2}}. Working and counter electrodes were made of graphite with a diameter of \SI{5}{mm}. Working sense and reference electrode were made from a 50~$\mu$m platinum wire. A \SI{1}{\mol/L} sodium chloride solution was used as electrolyte. We used a Gamry Instruments Reference 3000 potentiostat and applied a sinusoidal alternating current at a potential of \SI{10}{mV} with frequencies ranging from 300 kHz to 0.1 Hz.
Before each experimental cycle a specific amount of electrolyte was provided, to be used for all following measurements. Its inherent resistance and behavior in the test cell was determined by running a first measurement without the application of a membrane to the test cell. Each test run was repeated three times to verify consistency of the results. 

\subsection{Operation of FCDI Experiments}
\label{sec:runFCDI}

All experiments were performed with \SI{5}{g/L} sodium chloride water feed solutions. The solutions were prepared using ultrapure water and sodium chloride (Carl Roth GmbH \& Co. KG, 99.8~$\%$). The water feed streams were pumped by peristaltic pumps (Ismatec Reglo ICC peristaltic pump with up to three independent channels) setting the feed flow rates to \SI{2}{mL/min} each. Conductivity analysis of both, diluate and concentrate streams, were preformed with conductivity probes (Knick SE 615/1-MS, Knick Elektronische Messger\"ate GmbH \& Co. KG).
For the flow electrodes, activated carbon powder (Carbopal SC11PG, Donau carbon GmbH) was suspended in high-purity water with a mass fraction of 20~wt\% and left to stir for at least 24 hours. The flow electrodes were pumped by peristaltic pumps (Cole Parmer Masterflex Easyload II, Tube No.16) and the flow rate was set to \SI{200}{mL/min} for all experiments. For each flow electrode, \SI{100}{mL} activated carbon suspension was given into a glass vessel, where it was stirred with agitators to prevent settling of the activated carbon particles and pumped circularly through the modules back to the glass vessels. 

Teflon and polyethylene tubes with an inner diameter of \SI{4}{mm} connected the vessel with the used modules and the modules among each other. A potential of \SI{1.2}{V} was provided to all modules by power supplies which also measured the responding current (HM8143, Rohde \& Schwarz GmbH \& Co. KG, E3640A, Keysight Technologies, Inc.).

\begin{figure}[h!]
	\centering
		\includegraphics[width=90mm]{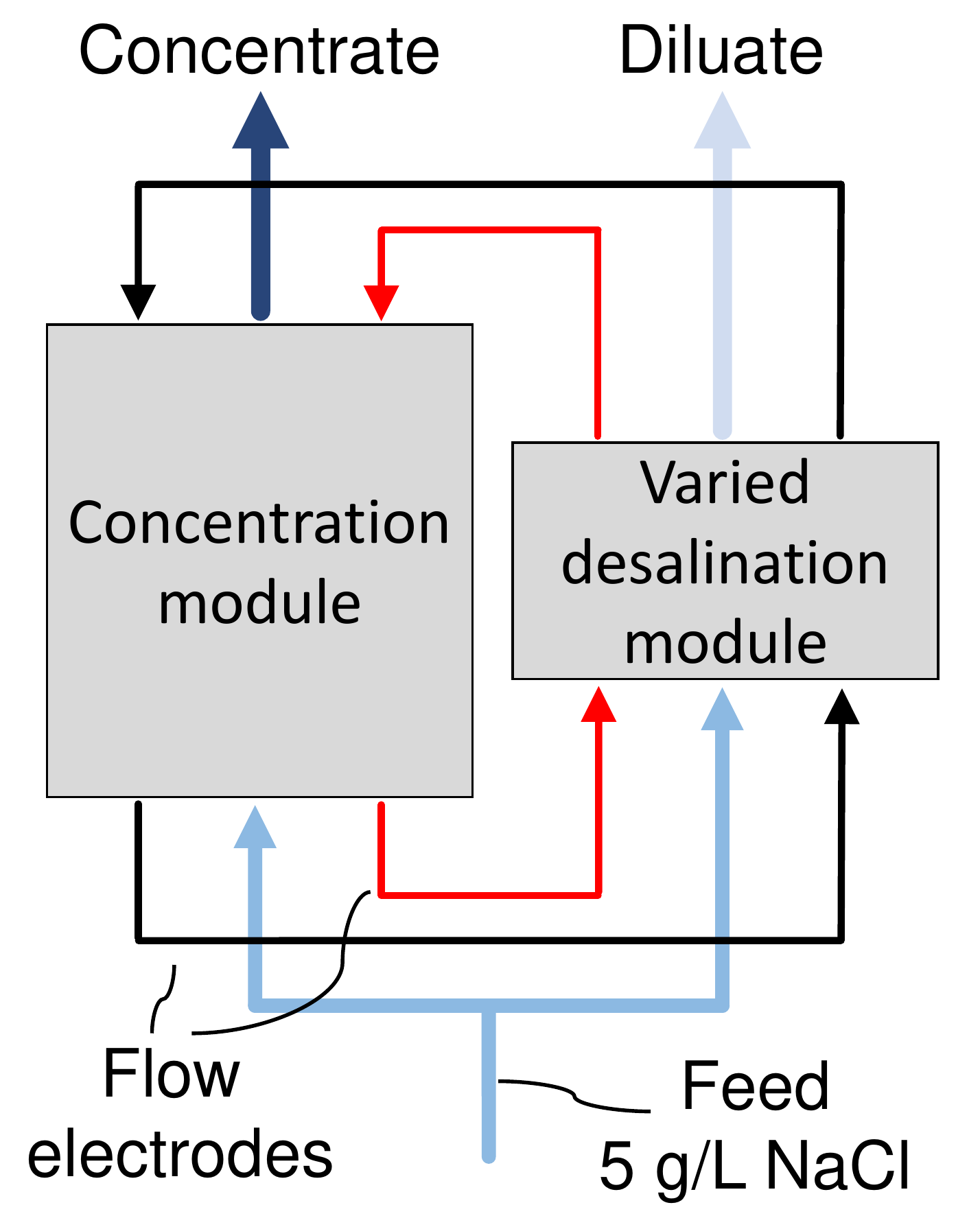}
	\caption{Module setup of two FCDI Modules for comparing Experiments. The left module for concentration has an active membrane area of 100~$cm^2$. The right desalination module was mounted with MEAs or graphite electrodes.}
	\label{fig:ModuleSetupConti}
\end{figure}

All experiments were operated continuously with two modules as presented in Figure~\ref{fig:ModuleSetupConti} until a steady state was established. Steady state was usually reached within 90 minutes. The key intention was to examine the behavior of the different membrane electrode assemblies in contrast to a graphite current collector. The right module represents the 'small' desalination module equipped with either graphite current collectors or MEAs, whereas the left module displays the 'standard' module used for regeneration of the flow electrodes. A larger module was chosen to avoid a possible bottleneck during regeneration of the flow electrodes, since the desalination performance was subject of examination. The flow electrodes were circulated between both modules, allowing a continuous operation. As feed, a sodium chloride solution was given into both modules. In order to counter the pressure loss in the module and fouling of the tubing, the flow rate was adjusted before every procedure and checked during operation. The conductivities of the water streams were recorded and used for further calculations to determine salt concentrations. A calibration curve was used to calculate the salt concentration from the measured conductivity. The pH value was checked in samples from the feed and diluate solution. The diagrams in the results section give the concentrations normalized for the feed water concentration and the membrane area-related salt transport rate calculated by Equation \ref{eqn:salttransport}. 

\begin{equation}\label{eqn:salttransport}
    \mathrm{\dot m_{salt}^{''}} = \frac{\left ( c_{desal} - c_{feed} \right ) \cdot \dot V_{feed}}{A_{active}}
\end{equation}

\section{Results and Discussion}

In electrochemical applications, membrane electrode assemblies (MEAs) are well established. The present study is the first on introducing this technology to flow-electrode capacitive deionization. We produced specialized membrane assemblies and tested different components for the MEAs. These MEAs were evaluated with the key parameters used for electrochemical membrane characterization. Desalination experiments were conducted to compare the new product with currently used graphite current collectors.

\subsection{MEA production}

The membrane electrode assemblies were composed from ion-exchange membranes and different carbon fiber fabrics, as described in Section \ref{sec:MEAmanufac}. During the hot-pressing procedure a homogeneous distribution of heat and surface load is necessary to achieve uniform MEAs. Pre-studies showed that the best results were achieved with the setup and hot-pressing procedure described in this section. 

We used three different carbon fiber fabrics. All fabrics have no surface coating to prevent undesired electrical resistances. The three fabrics differ in thickness and width of the roving. Small samples were produced for each combination of carbon fiber fabric and anion- or cation-exchange membrane, respectively.  

\subsection{MEA swelling behavior}

Swelling is a major issue in flat-sheet membrane modules. The size change of the polymer material is usually limited by the sealing of such modules, which can lead to the formation of wrinkles in the membranes. Especially unreinforced membranes have a high tendency to wrinkle. These undulations in the membrane may cause blockage of the flow-channels next to the membrane. Hence, we used reinforced membranes in our previous studies and compare their behavior with the newly manufactured MEAs.

\begin{figure}[h!]
\begin{center}
       \includegraphics[width=90mm]{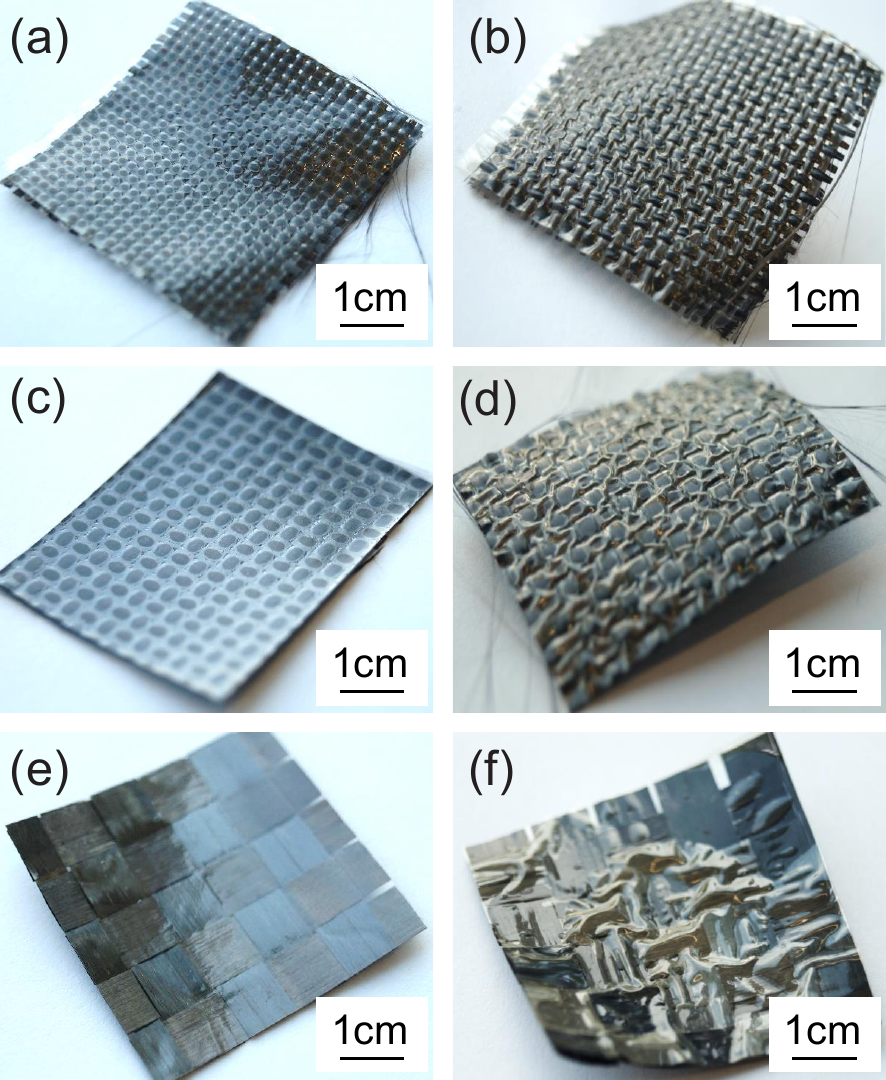}
\caption{Photographs of anion-exchange membrane-electrode assemblies manufactured from three different carbon fiber fabrics; (a) AMEA with CF92, (c) AMEA with CF200, (e) AMEA with CF46 and the same samples after swelling in water for 30 days; (b) swollen AMEA with CF92, (d) swollen AMEA with CF200, (f) swollen AMEA with CF46}
\label{fig:divMEAs}
\end{center}
\end{figure}

Figure~\ref{fig:divMEAs} displays the MEAs made from anion-exchange membranes in combination with carbon fiber fabrics with different surface weights. The pictures for MEAs made from cation-exchange membranes are displayed in the supplementary Figure~S4. As previously described in Section~\ref{sec:MEAmanufac}, different carbon fiber fabrics were used to create anion and cation-exchange membrane electrode assemblies (AMEAs/CMEAs) via hot pressing. Each carbon fiber fabric (CF) was successfully combined with an anion- (FAS~30) and cation- (FKS~30) exchange membrane. 

Sub-figures~\ref{fig:divMEAs}a, \ref{fig:divMEAs}c and \ref{fig:divMEAs}e display the AMEAs after hot-pressing. Both layers of the AMEAs are joined and cannot be detached. AMEA-CF46 shows no significant sign that it is pressed together with an IEM. The other two carbon fiber fabrics, however have distinct round patches where the IEM is combined with the carbon fiber fabric. In areas of roving crossing points the surface load has a maximum, allowing the IEM to bond with the carbon fiber fabric and forming the patches. The dimensions of the roving cross sections differ, also resulting in unequal patch sizes. CF46 has a cross section area of \SI{0.5}{cm^2}, CF92 has \SI{0.1}{cm^2}, whereas, CF200 has \SI{0.2}{cm^2}. Between the patches, the membranes fold themselves into the free space between the rovings creating grooves. This effect is different for each membrane material and is displayed in Figure~S3 of the supplementary material.

In Figures~\ref{fig:divMEAs}b, \ref{fig:divMEAs}d and \ref{fig:divMEAs}f, the AMEAs are shown after 30 days of storage in deionized water. Differences between the fresh MEAs and the the swollen MEAs become apparent. The membranes have many wrinkles all around the surface of the carbon fiber fabric. For AMEA-CF46 these wrinkles become so large that they cause detachment of the membrane from the carbon fiber fabric. The other two AMEAs also have many wrinkles, but only in the area where the rovings do not cross. The round patches are still visible. The interconnections leave space to let the membrane swell at these points. Similar occurrences can also be seen with CMEAs in Figure~S4 of the supplementary material. Attempts to improve the connection for AMEA-CF46 by tuning the hot-pressing parameters were not successful, no stable connection using CF46 could be achieved. In contrast, the connections in AMEA-CF92 and AMEA-CF200 are very durable and support the membrane. In these cases the carbon fiber fabric also acts as reinforcement of the membranes.

\subsection{MEA IEC}

Temperature control during the hot-pressing of the MEAs is essential, since the active groups of ion-exchange membranes could be destroyed. Thus, we set the temperature to \SI{120}{\celsius} to stay always below the critical range of \SI{160}{\celsius} and above. The critical value is a threshold from which desulfonation of sulfonated active groups inside of sulfonated polyether ether ketone (SPEEK) cation-exchange membranes (CEMs), such as the Fumatech FKS~30 applied in this study, starts \cite{Koziara2016}. For the anion-exchange membranes, no critical temperatures could be identified in the manufacturers data sheet or literature, so empirical analysis was necessary. For this purpose, the ion-exchange capacity (IEC) of the membranes used for MEA fabrication was determined before and after the hot pressing procedure at a chosen temperature. The following Table \ref{tab:IEC} shows the determined values in [mequiv/g].

\begin{table}[h!]
\centering
\vspace{2ex}
\caption[\textit{Results of ion-exchange capacity analysis}]{\textit{Results of ion-exchange capacity analysis in [mequiv/g].}}
\begin{tabular}{lcccl}
\cline{1-4}
\textbf{\begin{tabular}[c]{@{}l@{}}Type of \\ membrane\end{tabular}} & \textbf{\begin{tabular}[c]{@{}c@{}}Manufacturer\\   data\end{tabular}} & \textbf{\begin{tabular}[c]{@{}c@{}}Untreated\\   sample\end{tabular}} & \textbf{\begin{tabular}[c]{@{}c@{}}Hotpressed\\ at \SI{120}{C}\end{tabular}} &  \\ \cline{1-4}
FKS~30                                                               & 1.3-1.4                                                                & 1.36                                                                  & 1.37                                                                                &  \\
FAS~30                                                               & 1.6-1.8                                                                & 0.85                                                                  & 0.87                                                                                &  \\ \cline{1-4}
\end{tabular}
\label{tab:IEC}
\end{table}

No changes in IEC were observed for the unsupported thin film IEMs (FKS~30, FAS~30), which were used for the MEA fabrication. However, the IEC values of the FAS~30 membranes determined in this study differ significantly from the manufacturers data. This mismatch can be explained by the analytical procedure applied here, which is only capable of determining the amount of strong basic groups of the membrane material. It is likely that the FAS~30 anion-exchange membrane by Fumatech consists of strong basic as well as weak basic groups. Dlugolecki et al. discovered this for the Fumatech FAD anion-exchange membrane. To verify if weak basic groups are altered or destroyed during hot pressing, the application of other analytical methods would be required, which would go beyond the scope of this work \cite{Dugoecki2008}.

\subsection{MEA EIS}

\begin{figure}[h!]
\begin{center}
       \includegraphics[width=0.45\textwidth]{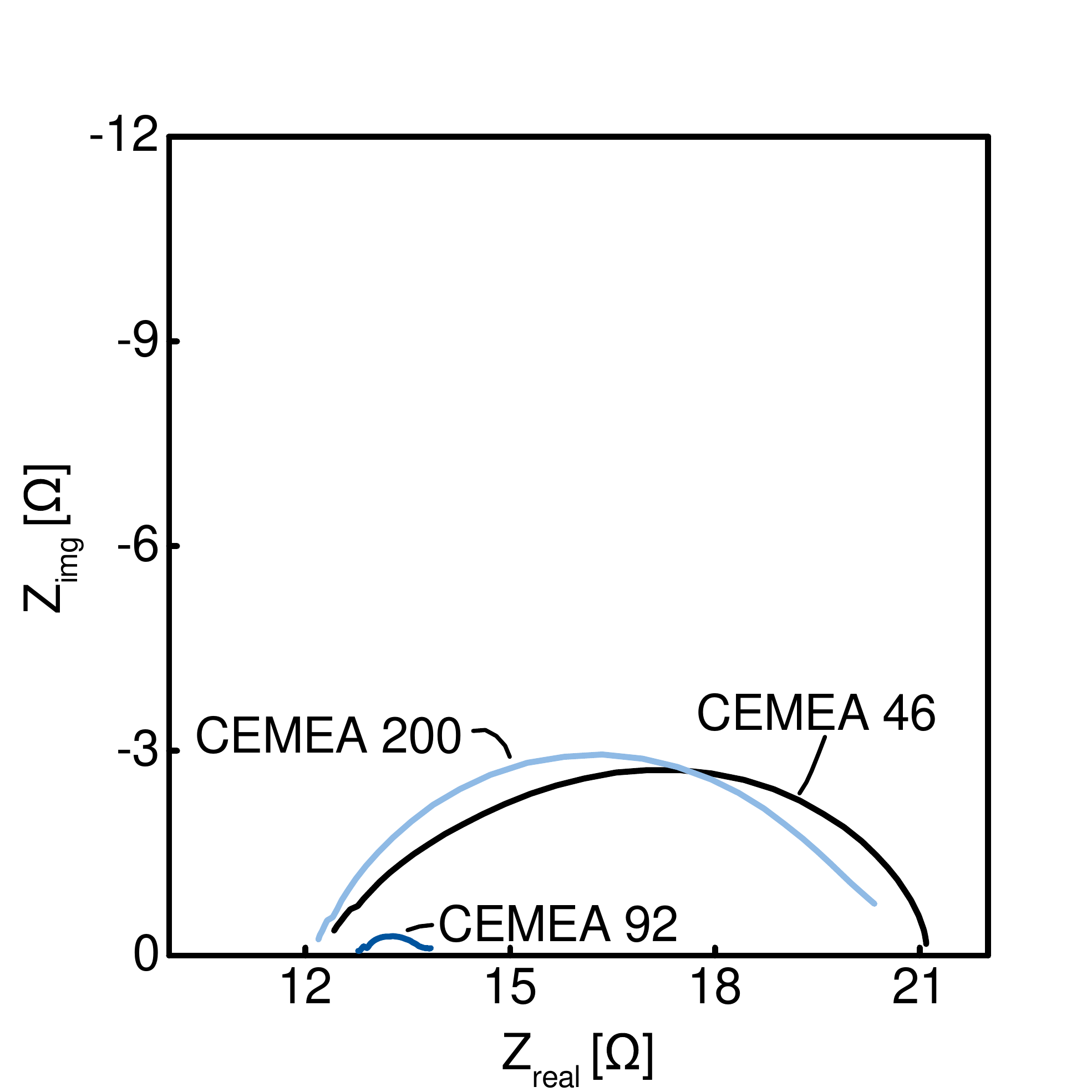}
\caption{Nyquist plot of the hot-pressed cation-exchange MEAs.}
\label{fig:R_EIS_CMEA}
\end{center}
\end{figure}

Figure~\ref{fig:R_EIS_CMEA} shows electrical impedance spectroscopy (EIS) measurement results of the fabricated MEAs in a Nyquist plot. The left boundary of all three semi-circles refers to a solution resistance of around 12~$\Omega$. This value was also measured as background resistance in the same measuring cell without a MEA installed, or with the fiber fabrics only. The width of the semi circles represents the double layer resistance and capacity referring to a simple Randles equivalent circuit. In this description, the measured double layer effects are a sum of the membrane and the carbon fiber fabric. Figure~\ref{fig:R_EIS_CMEA} displays that the MEA double layer resistance changes significantly depending on the type of carbon fiber fabric. The membrane resistance is equal, since the same membrane type (FKS~30) was used for the MEAs and thus, differences are caused by the different carbon fiber fabrics. It appears in this comparison that the MEA fabricated with \SI{92}{g/m^2} has the smallest resistance. A reason can be the tight weave of the carbon fibers and the overall thin fabric, which could leave paths for the ions to pass through easier. In addition, the CMEA fabricated with \SI{200}{g/m^2} could show comparable results, however, the thickness of this carbon fiber fabric is doubled compared to the carbon fiber fabric with \SI{92}{g/m^2}, revealing a greater overall resistance. In case of the carbon fiber fabric with \SI{46}{g/m^2}, the weave is flatter with a smaller thickness of 60~$\mu$m. This fabric has much more threads than the other two carbon fiber fabrics, which could lead to a higher resistance with both, anion-exchange and cation-exchange membranes. In summary, the total resistance depends on the interplay between density of the threads of the weaving and the thickness of the fabric. The carbon fiber fabric with \SI{92}{g/m^2} shows the overall best performance. Hence, we applied this carbon fiber fabric for the following desalination tests. The equivalent circuit used for the EIS analysis and the results for the AMEAs can be found in the supplementary Figure~S1.

\subsection{MEA FCDI Module Implementation}

\begin{figure}[h!]
\begin{center}
       \includegraphics[width=90mm]{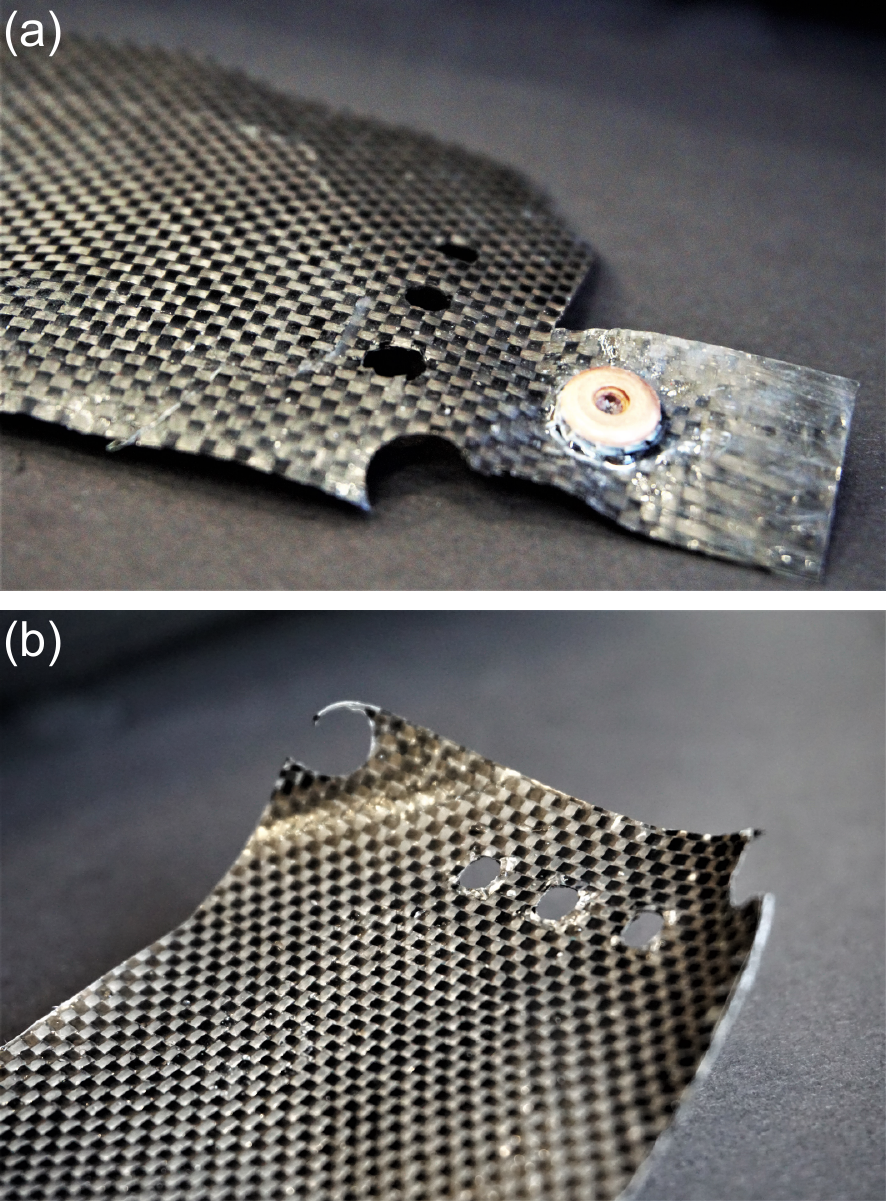}
\caption{Cation-exchange membrane electrode assembly for implementation in Fumatech ED-40 sized modules. (a) Fiber-side view with electrical contact and PDMS sealant. (b) Membrane-side view.}
\label{fig:SEMMEA}
\end{center}
\end{figure}

After testing different MEA configurations, we fabricated MEAs from the CF92 carbon fiber fabric and Fumatech FAS~30 and FKS~30 membranes in Fumatech ED-40 module size. The different MEAs were made as described in Section~\ref{sec:MEAmanufac}. Figure~\ref{fig:SEMMEA}a shows the side facing the flow electrode with the open carbon fiber fabric in the center. The rim is sealed with PDMS. Three smaller holes are punched through the MEA for distribution of both, flow electrodes and the salt water within the module. The holes along the sides are intended for the screws which bolt the module. The extension with the copper pop rivet can be seen on the right hand side of the image. The membrane side of the MEA is shown in \ref{fig:SEMMEA}b, where the holes have the same purposes as mentioned before. These MEAs were successfully implemented in \mbox{Fumatech ED-40} size FCDI modules.

\subsection{Desalination experiments}

\begin{figure}[h!]
\begin{center}
       \includegraphics[width=\textwidth]{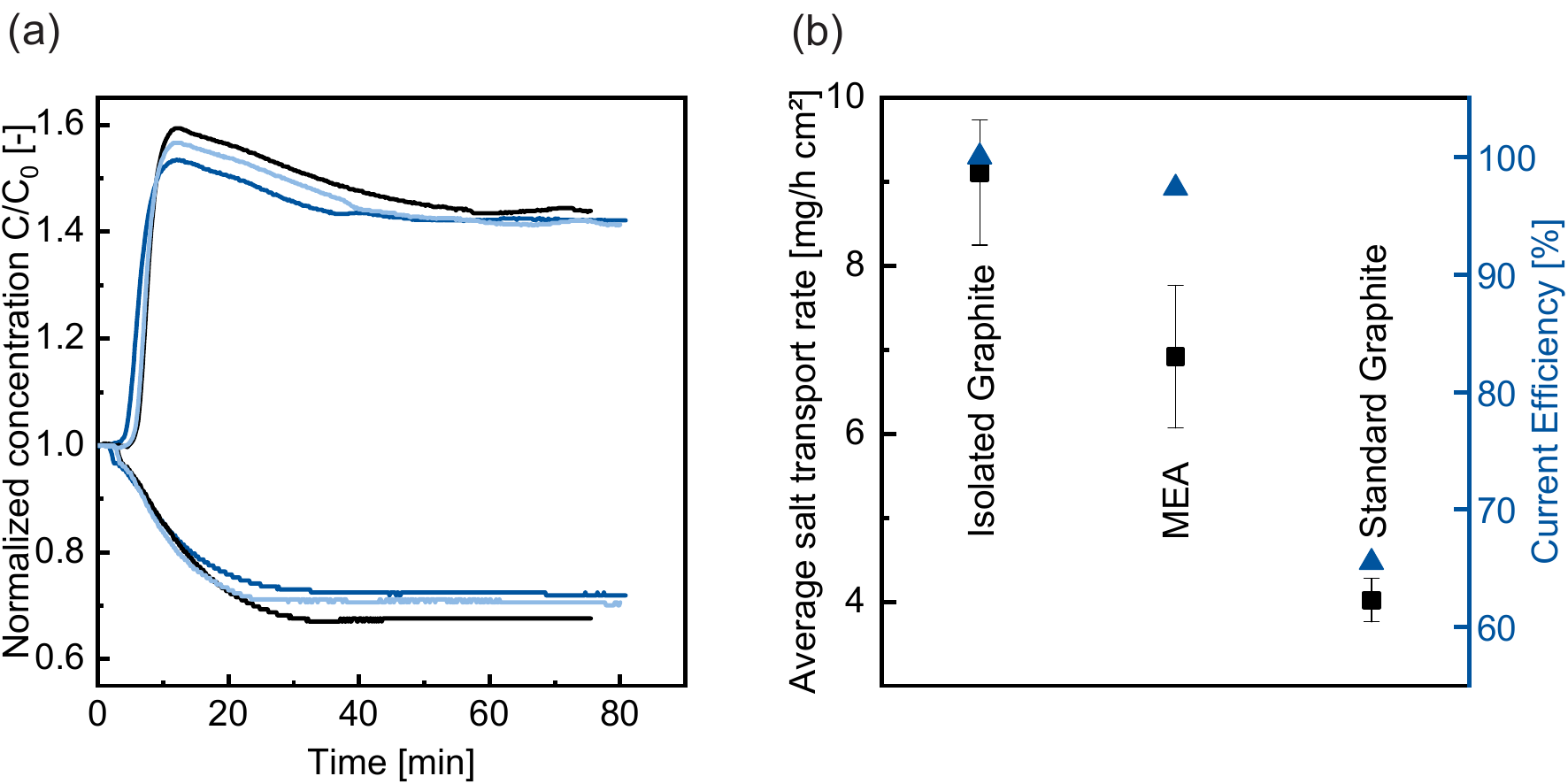}
\caption{Comparing FCDI experiments of a standard graphite FCDI module for concentration and different module setups for desalination. Liquid flow rates: 2mL/min; Slurry flow rates 200 mL/min; Potential at all modules: 1.2~V. (a) Course of normalized concentration over time showing the good comparability of three desalination experiments with isolated graphite current collectors. (b) Comparison of salt removal rate in steady state.}
\label{fig:moduleMEA}
\end{center}
\end{figure}

Desalination experiments were performed to examine the performance of the MEAs in an FCDI setup. The performance of the MEA experiments was compared to the performance of experiments with the isolated and non-isolated graphite current collectors shown in Figure~\ref{fig:ModuleComparison}. Figure~\ref{fig:moduleMEA}a displays the performance over time for the isolated graphite electrode as example. A concentrate and a diluate are produced during the experiments, which can be seen by an increase or decrease in the normalized concentration, respectively. Both curves reach a plateau, indicating that the overall system reached its steady state. Due to the fact that the module used for concentration offers 5.4 times more membrane surface area than the desalination module, the steady state is limited by the desalination speed only. Water samples were taken at the end of each experiment and the concentrations of both product streams were calculated from the conductivity and validated via HPLC. The pH of the samples shifted from pH 7.1 in the feed solution to pH 7.8-8.4 in the diluate and concentrate. With the flow rate of the streams and the active surface area of the desalination module of  \SI{22.4}{cm^2} the salt transport balances were calculated for desalination and concentration module of each experiment. Both balances, calculated via concentrate and diluate, showed equal results and the arithmetic mean was used for comparison. Figure \ref{fig:moduleMEA}b displays the comparison of the calculated salt transport and current efficiencies by Faraday's law. We tested three different setups, (a) the standard graphite current collector, (b) the MEAs, and (c) the same standard graphite current collector with isolated bore holes. In setup (c), the slurry can only exchange charge within the flow field facing the membrane. All setups were designed such that they can be supplied with the flow electrodes and feed water from the same side. This design was chosen to allow an easier scale-up in the future.  Each configuration of graphite collectors or MEAs were tested at least three times. It is clearly visible that the MEA setup offers around 175\% times more salt transport compared to the standard graphite, but also around 30\% less than the isolated graphite. This result is remarkable because (1) the active surface area for charge exchange of the MEA is only \SI{1274}{mm^2} compared to the graphite setups \SI{3078}{mm^2}, and (2) the effect of isolating the bore holes in graphite current collectors is significant. Additionally, both MEA setups and the isolated graphite setup show current efficiencies of almost 100\% compared to less than 70\% achieved with the standard setup. The potential for all experiments was set to 1.2~V and the mean current is in the range of 80~mA for the isolated graphite setup, 57~mA for the MEA setup and 58~mA for the standard graphite setup.

As shown in Figure~\ref{fig:ModuleComparison}, the anodic
flow electrode has to pass through the cathode current collector. Apparently the small contact area in the bore hole is enough to interfere with the charging at the anode. This leads to the reduced salt adsorption within that module. This result should be discussed in future work in more detail and needs to be taken into account when designing an FCDI module. Compared to the standard setup, the MEA setup achieves higher salt transfer rates and also adds beneficial properties of being flexible and thin.  
 
Regarding (1), the applied flow-field has a rectangular cross-section. Thus, flow electrode which is flowing through this flow-field carved into graphite is in contact with three conductive surfaces. Only the side facing the membrane surface is not conductive as displayed in Figure~\ref{fig:singlepoint}a. Contrary, when using the MEAs, only the surface facing the membrane is conductive but the experiments show a comparable desalination capacity. To describe this effect, the most natural explanation is that the slurry particles exchange both, electrical and ionic charges, starting from a single point of contact, as shown in Figure~\ref{fig:singlepoint}b. Charge percolation can still emerge forming percolation networks from the carbon fiber fabric towards the non-conductive flow-field. In total, these better and shorter charge transport paths leads to higher salt transport rates as standard graphite setups, even with less electrical charge exchange surface. This interpretation can be further proven through simulations as done by Lohaus et al. \cite{LOHAUS2019104}.

\section{Conclusion}

In electrochemical applications, membrane electrode assemblies (MEAs) are well established. In the present study, the technology is introduced to flow-electrode capacitive deionization. We tested different membranes and carbon fiber fabrics to produce the MEAs. These MEAs were evaluated with the key parameters used in the characterisation of membranes for electrochemical applications. Desalination experiments were conducted to compare the new product to the currently used graphite current collectors.

MEAs were successfully fabricated using a hot pressing procedure. The ion-exchange membranes swell and show many wrinkles in aqueous environment, but stay in good contact with the carbon fiber fabric. Thus, in addition to distributing the applied currents, the carbon fiber fabrics act as functional reinforcement. Depending on the thread density and thickness of the carbon fiber fabric, the swelling behavior differs. The same parameters of the carbon fiber fabric also have an influence on the MEA resistances obtained by electrical impedance spectroscopy. A thin and relatively open weaving is more favorable than thick and dense structures.

In desalination experiments, the performance of the most suitable MEA type was compared with our standard FCDI setup with graphite current collectors. When using graphite current collectors, the results differ significantly if the flow-electrode passing through the graphite current collector is electrically isolated from it. Since the MEAs are very thin, both flow-electrodes can enter the module unilaterally, without considerable losses in desalination or concentration performance. The MEA setup achieves salt transport rates comparable to the ones obtained with the two graphite current collector setups. The MEA setup performs better than the setup using non-insulated graphite plates, but not as good as the setup with insulated graphite current collectors. 

The current collector surface available for charge transfer is almost three times smaller in case of MEAs compared to the available charge transfer surface of the applied graphite current collectors. Thus, the transport mechanisms differ. We propose a new concept of an aligned charge transfer from a single point of contact, which combines shorter charge transfer paths and, hence reduces overpotentials. 

When using the MEAs in FCDI, there is no need for a conductive flow field anymore. Combined with the inherent flexibility of the MEAs, this creates a solid base for upscaling of FCDI systems in the future. 


\pagebreak
 \section*{Acknowledgement}
This work was supported by the German Federal Ministry of Education and Research (BMBF) under the project “ElektroWirbel” (FKZ 13XP5008) and by the European Research Council (ERC) under the European Unions Horizon 2020 research and innovation program (694946). The authors thank Fumatech BWT GmbH and Donau Carbon GmbH for the provision of material samples. The authors thank Karin Faensen for scanning electron microscope and microtomography images. The authors thank Tao Luo for his valuable input on ion-exchange membrane characteristics. The authors thank Zhaowei Zou and Georg Gert for their support and constructive effort.

 \section*{References}
\bibliographystyle{elsarticle-num}
\bibliography{Publication}{}

\end{document}